\documentclass{ws-ijmpe}

\begin{document}

\markboth{Wei-Liang Qian}{Effect of chemical freeze out on identified particle spectra at 200AGeV Au-Au Collisions at RHIC using SPheRIO}

\catchline{}{}{}{}{}

\title{Effect of chemical freeze out on identified particle spectra at 200AGeV Au-Au Collisions at RHIC using SPheRIO}

\author{\footnotesize Wei-Liang Qian
\footnote{Permanent address: Department of Physics, Fudan University, Shanghai, P.R. China, wlqian@fudan.edu.cn}, Rone Andrade, Fr\'{e}d\'{e}rique Grassi}
\address{
Instituto de F\'{\i}sica, Universidade de S\~ao Paulo\\
S\~ao Paulo, SP, C.P. 66318, 05315-970, Brazil\\}

\author{Otavio Socolowski Jr.}
\address{
Departamento de F\'{\i}sica, Instituto Tecnol\'{o}gico da  Aeron\'{a}utica - CTA\\
Pra\c{c}a Marechal Eduardo Gomes 50\\
S\~ao Jos\'e dos Camos, SP, C.P. 12228-900, Brazil\\}

\author{Takeshi Kodama}
\address{
Instituto de F\'{\i}sica, Universidade Federal do Rio de Janeiro\\
Rio de Janeiro, C.P. 68528, 21945-970, Brazil}

\author{Yogiro Hama}
\address{
Instituto de F\'{\i}sica, Universidade de S\~ao Paulo\\
S\~ao Paulo, SP, C.P. 66318, 05315-970, Brazil}

\maketitle

\begin{history}
\received{(received date)}
\revised{(revised date)}
\end{history}

\begin{abstract}
We investigate the effect of chemical freeze-out on identified particle spectra 
at 200AGeV Au-Au Collisions at RHIC, by utilizing a full three-dimensional hydrodynamical calculation.
The hydrodynamical code SPheRIO we employed is based on the smoothed particle hydrodynamic algorithm.
In order to describe the spectra of strange hadrons, the code has been further improved by explicitly incorporating the
strangeness conservation and a chemical freeze-out mechanism.
In our model, strange hadrons such as $\Lambda$, $\Xi$, $\Omega$ and $\phi$ 
undergo the chemical freeze-out immediately after the hadronization,
and their multiplicities are fixed thereafter.
At a lower temperature the thermal freeze-out takes place for all the particles.
It is shown that the present model provides a reasonably good description for the
spectra of identified particles, in particular, considerable improvement is observed for those of strange hadrons.
\end{abstract}

\section{Smoothed particle hydrodynamic model}
Statistical model and hydro-inspired model calculations indicate the picture of early chemical freeze-out(CFO)
for strange hadrons
in relativistic heavy ion collisions\cite{ch0,ch1,ch2,ch3,ch4,ch5}.
The CFO temperature can be extracted by
fitting the model calculation of particle ratios for hadrons to the experimental data.
On the other hand, one obtains the thermal freeze-out temperature from the
slope of transverse momentum distributions by assuming some radial flow profile.
Naturally, it is therefore motivating to see how this scenario can be adopted 
by more realistic models,
where a full three-dimensional hydrodynamical evolution is taken into account.
In this work, we present results of hydrodynamical model calculations,
inspired by this picture.
The hydrodynamical model we employed is based on smoothed particle hydrodynamic(SPH)
algorithm\cite{topics,va},
In this model, the matter flow is parametrized in terms of discrete Lagrangian coordinates,
of the so-called SPH particles. As a result, the hydrodynamic equations are reduced to
a system of coupled ordinary differential equations.
The code which implements the entropy representation of the SPH model for
relativistic high energy collisions, and which has been developed within the 
S\~{a}o Paulo - Rio de Janeiro Collaboration, is
called SPheRIO. As it has been shown, the model is an efficient and robust method to 
tackle the problems concerning relativistic high-energy nucleus-nucleus collisions, 
which are characterized by highly asymmetrical configurations.
It has been successfully used to investigate the effects of the initial-condition fluctuations and adopting
the continuous emission scenario for the description of decoupling process 
\cite{va,topics,ce,ebe,strange,v2}.

In the present calculation, the model has been further improved in order to consider
strangeness conservation and to adopt the scenario of CFO.
The strangeness conservation is implemented by explicitly incorporating
strangeness chemical potential into the code, and correspondingly a different set of
equation of state(EOS) has been built and utilized.
As a good approximation, we assume local strangeness neutrality throughout the hydro evolution.

To adopt the picture of CFO, as a first step,
we take a simple postulate that for strange hadrons such as $\Lambda$, $\Xi$, $\Omega$ and $\phi$,
the CFO takes place immediately after they are completely hadronized.
These particles cease to have inelastic collisions and therefore their abundances are 
fixed and determined only by partonic EOS.
Instead of introducing a specific temperature,
CFO is incorporated in a simple and parameter-free approach.
This is inspired by the earlier estimation from statistical model\cite{ch1}
that CFO takes place close to the phase transition region.
It is also because at this point, we would like first to check
whether the picture of CFO would qualitatively improve the result of hydrodynamic model
without going into the details of parameters.
With the system being cooled and rarefied further,
the thermal freeze-out occurs at a lower temperature $T_f $
which serves as an adjustable parameter in our calculation.

\section{Results and discussions}

As in the previous calculations, we use NEXUS event generator to 
produce event-by-event fluctuating initial conditions.
We make use of a rescaling factor\cite{peter} to fix
the pseudo-rapidity distribution for all charged particles.
In the present calculation, we introduce
another parameter $\alpha$ which provides initial transverse
velocity in addition to the one provided by NEXUS.
The reason for the introduction of $\alpha$ is because NEXUS almost
does not provide any initial transverse velocity.
Thus the initial transverse expansion reads
\begin{eqnarray}
v_T &=& NEXUS + \tanh(\alpha r)\hat{r}
\end{eqnarray}
where $r$ is the radial distance from orgin.

\begin{figure}[!htb]
\vspace*{-1cm}
\centerline{\psfig{file=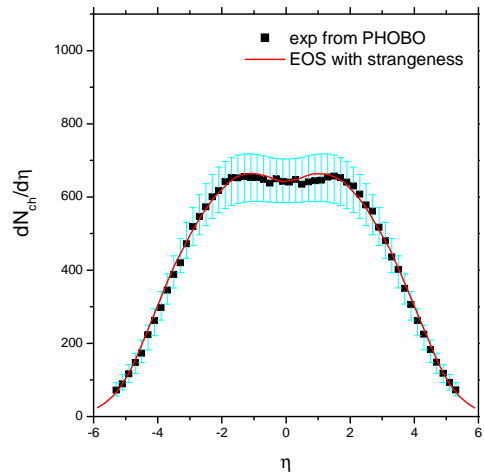,width=7.5cm}}
\vspace*{-3cm}
\label{figure:fig1}
\caption{The pseudo-rapidity distribution for all charged particles for the
most central Au+Au collision at 200A GeV. The data are from PHOBOS Collab.} 
\end{figure}

As a preliminary application, we have calculated the
pseudo-rapidity distribution for all charged particles. In Fig.1,
we present the pseudo-rapidity distribution for the most central Au+Au collisions
at 200A GeV. The experimental
data are from PHOBOS Collab, taken in the most central Au+Au at 200A GeV.

\begin{figure}[!htb]
\vspace*{-1cm}
\centerline{\psfig{file=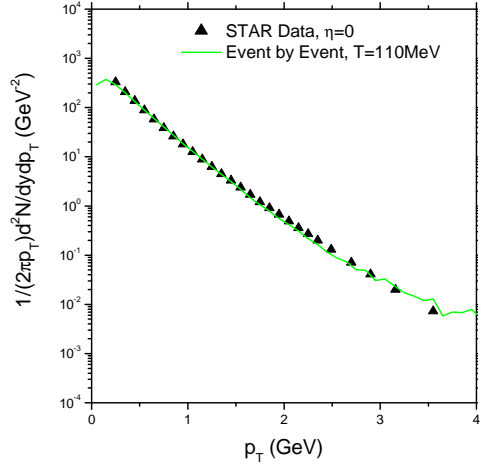,width=7.5cm}}
\vspace*{-3cm}
\label{figure:fig2}
\caption{Transverse momentum distribution of all charged particles for the
most central Au+Au collision at 200A GeV in the rapidity interval $-1.0 < \eta <
1.0$. The data are from STAR Collab.} 
\end{figure}

We show in Fig.2, the experimental transverse momentum distribution data for all charged
particles, which can be well reproduced with
a choice of $\alpha=0.04 $fm$^{-1}$ and freeze-out temperature $T_f = 110$ MeV.
The experimental data are from STAR Collab\cite{star1},
taken in the most central Au+Au at 200A GeV, with $\eta=0$.

\begin{figure}[!htb]
\vspace*{-1cm}
\centerline{\psfig{file=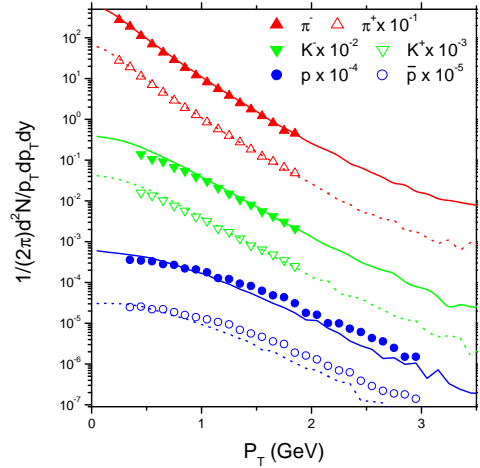,width=7.5cm}}
\vspace*{-3cm}
\label{figure:fig3}
\caption{Transverse momentum distribution of pions, kaons and protons for the
most central Au+Au collision at 200A GeV in the rapidity interval $-1.0 < \eta <
1.0$ for the most central collisions. The data are from BRAHMS Collab.} 
\end{figure}

\begin{figure}[!htb]
\vspace*{-1cm}
\centerline{\psfig{file=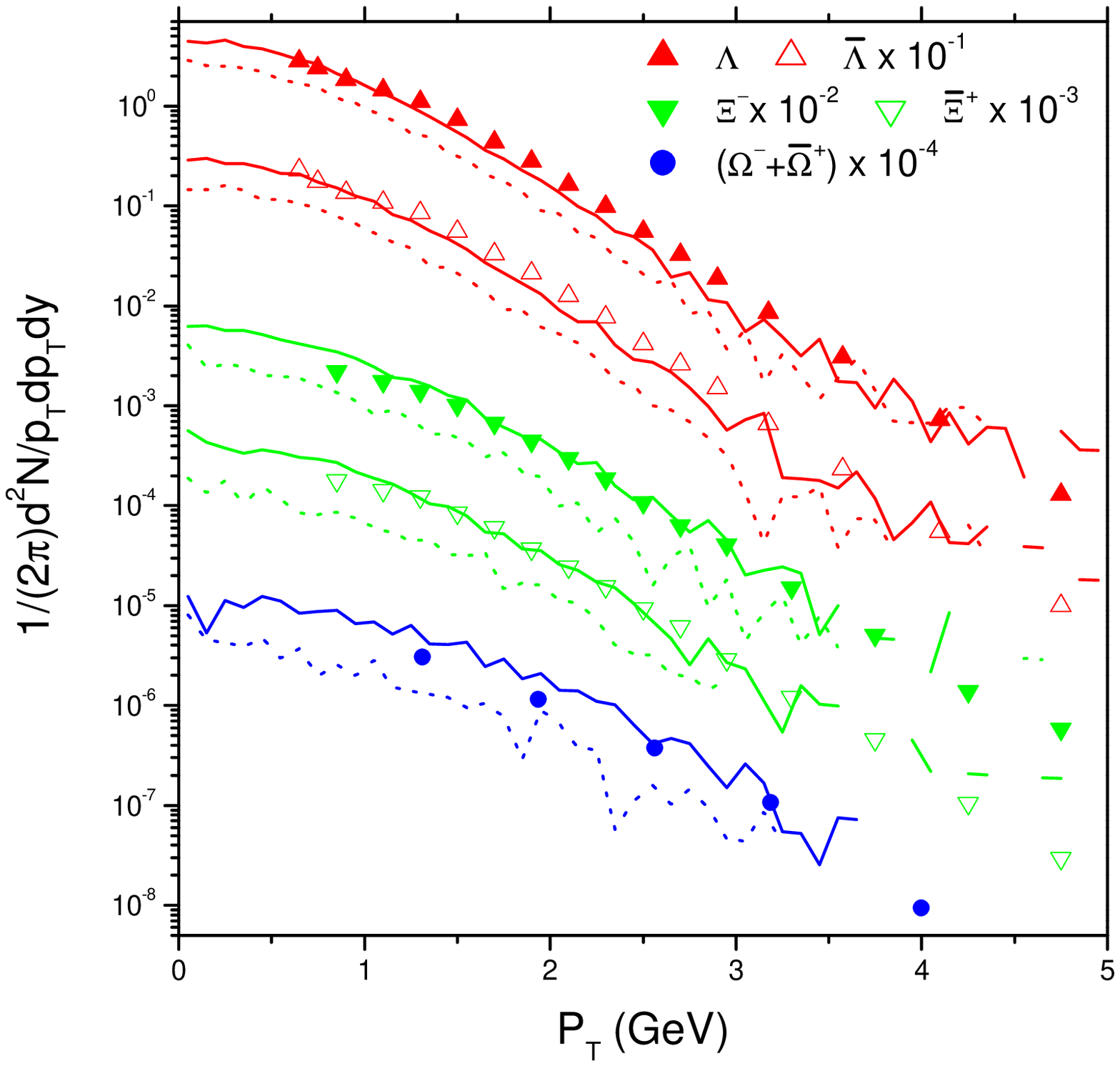,width=7.5cm}}
\vspace*{-3cm}
\label{figure:fig4}
\caption{Transverse momentum distribution of $\Lambda$, $\Xi$ and $\Omega$ for the
most central Au+Au collision at 200A GeV in the rapidity interval $-1.0 < \eta <
1.0$. The data are from STAR Collab. The dotted lines indicate
the results obtained without CFO, and solid line for
those with CFO incorporated in the model.} 
\end{figure}

With the parameters chosen, in the following, we calculate
the spectra for various hadrons.
In Fig.3 we show the transverse mass spectra
of pions, protons and kaons for most central collisions at mid rapidity, as well as
experimental data from BRAHMS Collab\cite{brahms1}.
The same spectra for $\Lambda$, $\Xi$ and $\Omega$ are depicted in Fig.4,
together with data from STAR Collab\cite{star2}. We use dotted lines to represent
the results obtained without incorporating CFO, and solid lines for
those with CFO switched on.

It is observed that the present hydrodynamic model gives good
description of the experimental transverse momentum spectra for
pions, kaons and protons even without turning on CFO for these particles. 
Although it gives the correct slopes 
for the spectra of strange hyperons such as $\Lambda$, $\Xi$ and $\Omega$,
the disagreement comes from the multiplicities of the spectra.
By introducing CFO, those results are improved significantly.
While the spectra for pions, kaons and protons almost remain the same, (this is not shown in the figures.)
it provides a good fit of strange hyperons $\Lambda$, $\Xi$ and $\Omega$.
The CFO amplifies the multiplicities of strange hadrons with respect to the ones in the original model 
owing to higher CFO temperature, as compared with the thermal freeze out one.
Meanwhile the slopes of the spectra remain unchanged.
It was indicated experimentally in Ref.\cite{kdis},
as one goes to large rapidity region where the baryon density differs sizably from zero, the outcome of CFO would be much
more significant. Further work on this topic is under progress.

It is worth noting that the effect of CFO has been discussed by several authors.
In ref\cite{hch1,hch2,hch3,hch4}, the early CFO is studied in term of hybrid model
where the hydodynamic evolution is complemented with a hadronic cascade
model. The pure hydrodynamic evolution calculation has been carried out either only
in the transverse plane assuming Bjorken's scaling\cite{hch5} or with the baryon chemical potential 
taken to be zero\cite{hch6}.

We acknowledge financial support by FAPESP (2004/10619-9, 2005/54595-9, 2004/13309-0 and
2004/15560-2), CAPES/PrOBRAL, CNPq, FAPERJ and PRONEX.

\end{document}